\documentclass[prl,twocolumn,floatfix,amssymb,showpacs]{revtex4}

\usepackage{graphicx}

\newcommand{\ud}{{\mathrm{d}}} 
\newcommand{\es}{\varepsilon_\mathrm{s}}
\newcommand{\eb}{\varepsilon_\mathrm{b}}
\newcommand{\dbar}{d_\mathrm{bar}}
\newcommand{\rA}{r_\mathrm{A}}

\begin{document}

\title{Gate-induced ionization of single dopant atoms}

\author{G.~D.~J.~Smit}
\email{g.d.j.smit@tnw.tudelft.nl}
\author{S.~Rogge}
\email{s.rogge@tnw.tudelft.nl}
\author{J.~Caro}
\author{T.~M.~Klapwijk}
\affiliation{Department of NanoScience, Delft University of
Technology, Lorentzweg 1, 2628 CJ Delft, The Netherlands}

\begin{abstract}
Gate-induced wave function manipulation of a single dopant atom is
a possible basis of atomic scale electronics. From this
perspective, we analyzed the effect of a small nearby gate on a
single dopant atom in a semiconductor up to field ionization. The
dopant is modelled as a hydrogen-like impurity and the
Schr\"odinger equation is solved by a variational method. We find
that---depending on the separation of the dopant and the
gate---the electron transfer is either gradual or abrupt, defining
two distinctive regimes for the gate-induced ionization process.
\end{abstract}

\date{\today}

\pacs{03.67.Lx, 85.30.De, 73.21.-b, 71.55.-i}

\maketitle


The size regime where the discreteness of doping must be taken
into account is brought within experimental reach by today's
semiconductor lithography techniques. In this regime, single
dopant atoms have been demonstrated to dominate the behavior of
downscaled versions of conventional devices \cite{calvet02}. On
the other hand, the promising opportunity is offered to study the
physics of semiconductors on their ultimate length scale by
addressing separate dopants. Putting a small gate close to a
single impurity would, for example, allow for the manipulation of
\emph{individual} hydrogen-like wave functions. Furthermore, large
electric fields (otherwise only achievable in astronomy) can be
experimentally obtained in semiconductors due to the occurrence of
large dielectric constants and small effective masses. Apart from
the fundamental importance, an ultimate application is found in a
Si-based solid state quantum computer \cite{kane98,kane00}, in
which the nuclear spins of single $^{31}$P-dopants are envisioned
as qubits. In this proposal, addressing a single qubit by NMR is
achieved via the hyperfine interaction of the nuclear spin and its
valence electron, which can be tuned by modifying the electron
wave function with a nearby gate. In a recent variation of this
design \cite{skinner03}, the ionization of single dopants by this
gate is an essential ingredient.

Our aim is to quantitatively investigate the effect of the
electric field generated by a local gate on a single neutral
dopant atom in a semiconductor, ultimately leading to ionization.
The response to small fields has been addressed before in the
context of quantum computing \cite{larionov00,pakes02}. In this
paper, the complete ionization process is discussed. Our approach
incorporates the computation of time independent ground state wave
functions of the system and, subsequently, the estimation of
transition probabilities. We conclude that the separation of the
dopant and the gate determines the nature of the ionization
process. When the dopant resides close to the gate, the electron
is gradually pulled away from the dopant when the gate voltage is
increased, while for a larger separation the dopant ionizes
abruptly at a well-defined gate voltage.

\begin{figure}
  \centering
  \includegraphics[width=8cm]{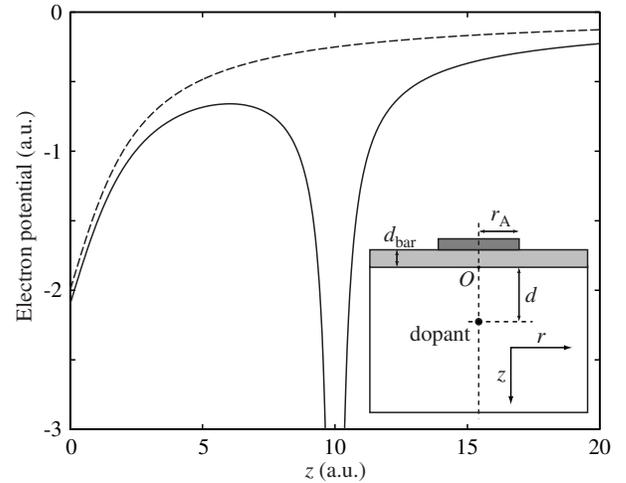}
  \caption{The dashed line represents the calculated potential
  due to the gate at the symmetry axis of the device for $\rA=2$~a.u.
  and a gate voltage of 2~a.u. The solid line includes the dopant
  potential for $d=10$~a.u. (Note that e.g.\ in silicon $1~\mathrm{a.u.}
  \approx 3$~nm for lengths and $1~\mathrm{a.u.}
  \approx 90$~mV for voltages.)
  The inset shows a schematic of the device layout, indicating the
  important parameters.}
  \label{fig:elpot}
\end{figure}

Addressing a single dopant requires a small local gate. When a
dopant would be ionized by a large gate (e.g. an infinite strip
\cite{pakes02}), the electron would be delocalized along the gate.
This would be undesirable in applications where (spin-)phase
coherence must be kept under control, such as a quantum computer.
Therefore, we chose to model the gate as a circular disc, having
the additional advantage that the complete system (dopant plus
gate) is radially symmetric. The layout of our model system is
schematically depicted in the inset of Fig.~\ref{fig:elpot}. The
disc-shaped metallic gate with radius $\rA$ is separated from the
semiconductor bulk (relative dielectric constant $\es$) by a
barrier (relative dielectric constant $\eb$) of thickness $\dbar$.
A dopant is positioned at distance $d$ from the
barrier-semiconductor interface and centered with respect to the
gate.

At low temperatures, the semiconductor can be considered as a
dielectric, due to the absence of free charges. Charges at the
barrier-semiconductor interfaces and in the barrier will be
neglected. In our calculations, we assume the barrier to be
infinitely high and infinitely thin ($\dbar = 0$), which allows us
to take advantage of the fact that the potential due to a charged
metallic disc in a uniform dielectric medium can be expressed in
closed form \cite{smythe50} (we will demonstrate the applicability
of our results to a realistic layout). The total potential was
obtained by adding a Coulomb potential well due to a positive unit
charge in the semiconductor. A cross-section of the total electron
potential for some typical parameters is shown in
Fig.~\ref{fig:elpot}. Image charge effects at the
semiconductor-barrier and the barrier-gate interfaces were
neglected.

In our calculations, the dopant atom is described within an
effective mass approach: the contribution of the semiconductor
bandstructure is accounted for by considering it as an uniform
dielectric medium and using an isotropic effective mass. Such a
hydrogen-like model is known to provide a good first order
description of a dopant atom (although it fails to accurately
describe the energy levels \cite{ramdas81} and interactions
\cite{koiller02}). It is sufficient for our purpose and allows us
to capture crucial phenomena and obtain estimates of important
parameters. To keep our results general and transparent, physical
quantities will be expressed in (effective) atomic units (a.u.)
\footnote{(Effective) atomic units comprise setting the reduced
Planck's constant $\hbar$, the electron charge $e$ and the
electron effective mass $m^*$ equal to unity. As a result, length
is expressed in units of the effective Bohr radius $a_0^*$ and
energy is expressed in units of twice the effective ionization
energy of the dopant (i.e. twice the effective Rydberg Ry$^*$).}.
To simplify the conversion to conventional units, some values for
silicon are given as an example in the caption of
Fig.~\ref{fig:elpot}.


The time-independent Hamiltonian of the problem reads (in atomic
units)
$$
  \mathcal{H}=-\frac{1}{2}\nabla^2-\frac{1}{\sqrt{r^2+(z-d)^2}}+
  V_\mathrm{g}(r,z),
$$
where $V_\mathrm{g}(r,z)$ describes the potential landscape in the
semiconductor due to the gate and $(r,z)$ are cylinder coordinates
as defined in Fig.~\ref{fig:elpot} (inset). Approximate ground
state wave functions are found by a variational method. As trial
wave function we use a linear combination of functions from a
fixed and finite set $\mathcal{S}$, where the weights are used as
variational parameters. To this end, we choose $\mathcal{S}$ to
contain functions of the form
\begin{equation}
  \varphi(r,z)=\exp(-\alpha r^2)\cdot z\exp(-\beta(z-d)^2)
  \label{eq:phidop}
\end{equation}
and
\begin{equation}
  \tilde{\varphi}(r,z)=\exp(-\gamma r^2)\cdot z\exp(-\delta z^2),
  \label{eq:phigate}
\end{equation}
where $\alpha$, $\beta$, $\gamma$, and $\delta$ are constants that
will be chosen later. The functions are cylinder-symmetric,
motivated by the radial symmetry of the potential and the fact
that the ground state is expected to be $s$-like. To allow for a
full description of the ionization process, it is important that
$\mathcal{S}$ includes both wave functions of the form
(\ref{eq:phidop}), having large electron density at the dopant
site, and of the form (\ref{eq:phigate}), where the electron
resides close to the gate.

The functional form of Eq.~(\ref{eq:phidop}) is motivated by the
fact that the (exponential) ground state wave function of
hydrogen-like atoms can be quite well approximated as a linear
combination of gaussians \cite{huzinaga65}, which are much easier
to work with numerically. To make sure that the wave functions
vanish at the interface ($z=0$), it is multiplied by $z$. The
$\varphi(r,z)$ are allowed to become aspherical due to the gate
action by choosing different values for $\alpha$ and $\beta$.
Concerning the form of Eq.~(\ref{eq:phigate}), we note that the
potential well caused by the gate can in the radial direction be
approximated by a parabola. Consequently, a ground state wave
function similar to that of a linear harmonic oscillator is
expected and therefore the $r$-dependent part of
$\tilde{\varphi}(r,z)$ is chosen as a gaussian. The ground state
wave function of the triangular shaped well in the $z$-direction
can be approximated as $z\cdot e^{-\zeta z}$
(Ref.~\onlinecite{fang66}). Again, we will approximate the
exponential by a linear combination of gaussians.

In order to choose concrete values for the constants $\alpha$,
$\beta$, $\gamma$, and $\delta$, we note that for each positive
integer $N$ it is possible to find a set of $N$ real numbers
$\{\lambda_i\}_{i=1}^N$, such that a linear combination of
$\exp(-\lambda_i r^2)$ optimally approximates the ground state
wave function of hydrogen \cite{huzinaga65}. We will use the
values given in Ref.~\onlinecite{huzinaga65}, which are, for
example, $\{0.101, 0.321, 1.15, 5.06, 33.6\}$ for $N=5$. In our
calculation, we created functions of type (\ref{eq:phidop}) by
taking values for $\alpha$ and $\beta$ from such a set in all
possible $N^2$ combinations. Functions of type (\ref{eq:phigate})
were created by choosing values for $\gamma$ and $\delta$ from the
same set, after multiplying all elements by the scaling constant
$r_\mathrm{A}^{-1/2}$ to account for the size of the gate.
Proceeding like this, $\mathcal{S}$ contains a total of $2N^2$
functions. It was found that taking $N>5$ did hardly improve the
accuracy. Therefore, $N=5$ was used in all presented results.

Denoting the elements of $\mathcal{S}$ by $\psi_n$, the
variational procedure is now performed by forming the trial wave
function
$$
  \psi(r,z)=\sum_{\psi_n\in\mathcal{S}}c_{n}\psi_{n}(r,z)
$$
as a linear combination of the $\psi_n$ and minimizing the
functional
$$
  \frac{\langle\psi|\mathcal{H}|\psi\rangle}{\langle\psi|\psi\rangle}
$$
with respect to the variational parameters $c_n$. This minimum is
an upper bound to the ground state energy of $\mathcal{H}$. This
variational problem is equivalent to finding the smallest
eigenvalue of the generalized matrix eigenvalue problem
\begin{equation}
  (\mathbf{H}-E\mathbf{M})\cdot \mathbf{c}=0,
  \label{eq:eig}
\end{equation}
where $\mathbf{H}$ is the hamiltonian matrix expanded on the
$\psi_n$ with elements
$H_{ij}=\langle\psi_i|\mathcal{H}|\psi_j\rangle$ and $\mathbf{M}$
is the overlap matrix of the $\psi_n$ defined as
$M_{ij}=\langle\psi_i|\psi_j\rangle$. Furthermore,
$\mathbf{c}=(c_1,c_2,\ldots,c_m)$ and the inner-product
$\langle\cdot|\cdot\rangle$ is (as usual) defined as
$$
  \langle\psi_i|\psi_j\rangle=
  \int_0^\infty\int_0^\infty \psi_i^*(r,z)\psi_j(r,z)
  2\pi r\,\ud r\ud z.
$$
Note that $\mathbf{M}$ would be the unit matrix if $\mathcal{S}$
would be an orthonormal set with respect to
$\langle\cdot|\cdot\rangle$. In that case, Eq.~(\ref{eq:eig})
would reduce to an ordinary eigenvalue problem.

The smallest eigenvalue $E_0$ of Eq.~(\ref{eq:eig}) is an upper
bound to the ground state energy of the system. When $\mathcal{S}$
is chosen properly, $E_0$ is a good approximation to the real
ground state energy of $\mathcal{H}$ and the corresponding
eigenvector $\mathbf{c}$ defines a wave function that is a good
approximation of the real ground state wave function.

\begin{figure}
  \centering
  \includegraphics[width=8.6cm]{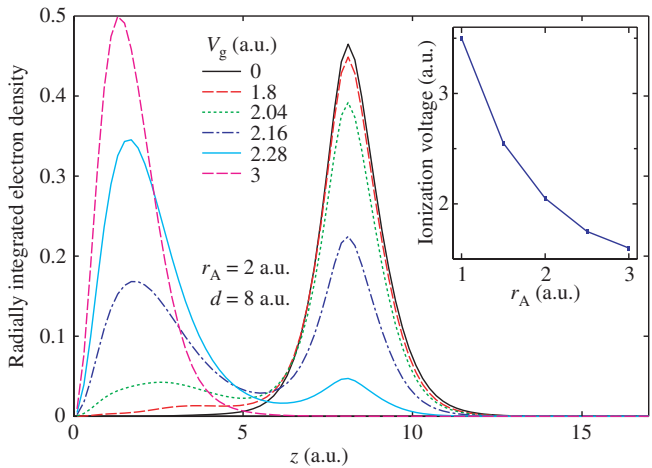}
  \caption{(Color online) The radially integrated probability
  density of the electron wave function as a function of $z$ for
  various gate voltages. The inset shows the ionization voltage versus
  gate radius $\rA$. (In silicon: $1~\mathrm{a.u.}
  \approx 3$~nm and 90~mV, respectively.)}
  \label{fig:dens}
\end{figure}

Once this wave function is known for several values of the dopant
depth $d$, gate voltage $V_\mathrm{g}$ and gate radius
$r_\mathrm{A}$, we will use it to study the ionization process of
the dopant. As an example, the radially integrated probability
density of the calculated electron wave function (i.e.
$\int_0^\infty |\psi(r,z)|^22\pi r\,\ud r$) is plotted versus $z$
in Fig.~\ref{fig:dens}. At zero gate voltage \footnote{Note that
zero gate voltage is defined as the flat-band situation in the
semiconductor. Due to interface effects at the gate (that result
in band bending), this might correspond to an actual gate voltage
that is non zero.}, the electron occupies the dopant site. For
increasing gate voltage, the electron is gradually pulled away
from the donor site. Finally, for large enough gate voltage, it
resides completely in the newly created potential well at the
gate.

\begin{figure}
  \centering
  \includegraphics[width=8.6cm]{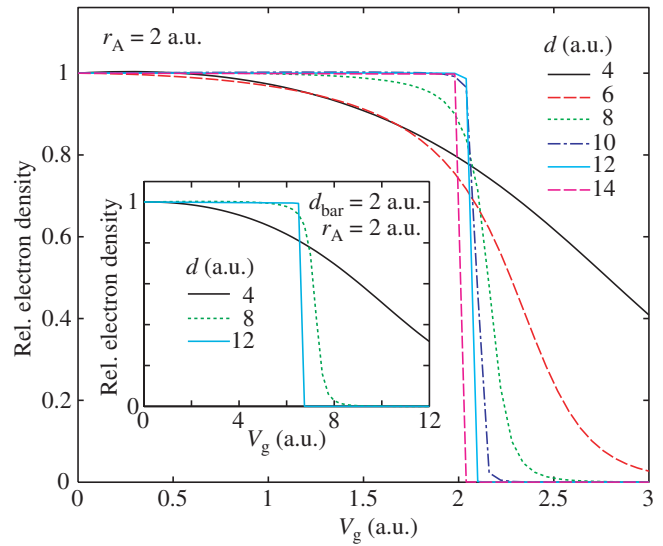}
  \caption{(Color online) The electron density at the dopant
  site as a function of gate voltage for various dopant depths,
  showing the process of ionization. All curves are
  normalized to their value at $V_\mathrm{g}=0$. The
  transition from a smooth to a step-like behavior is clearly
  visible at $d\approx 8a_0^*$.
  The inset shows the corresponding data with a 2~a.u. thick oxide
  barrier present. The behavior is similar, but occurs at
  higher gate voltage.}
  \label{fig:hyper}
\end{figure}

An interesting physical quantity is the electron density
$|\psi_0(0,d)|^2$ at the dopant site, as derived from the
approximated ground state wave function $\psi_0(r,z)$. We will use
it as an indication of the position of the electron: when the
electron is pulled away from the dopant site, this number
decreases. Moreover, it is of physical importance because the
hyperfine interaction is proportional to this number
\cite{bethe57}.

The characteristics of the electron transfer from the dopant to
the gate with increasing gate voltage depend on the distance $d$
of the dopant under the gate. In Fig.~\ref{fig:hyper}, the
electron density at the dopant site $|\psi_0(0,d)|^2$ (normalized
to the value at zero gate voltage) is plotted as a function of
gate voltage for several values of $d$. It can be seen that for
small $d$ the electron is transferred gradually from the dopant to
the gate, while for larger $d$ an abrupt electron jump occurs,
defining an ionization voltage. This can be explained from the
fact that for large $d$, a sufficiently large barrier separates
the two potential wells. For small $d$, the two wells are so
strongly coupled that they can be considered as a single well, the
position of which is pulled towards the gate with increasing gate
voltage.

The calculations were repeated for several gate radii $\rA$. The
inset of Fig.~\ref{fig:dens} shows the ionization voltage for
$d=15$ versus $\rA$. From the figure it is clear that the voltage
gets smaller for larger $\rA$. The reason for this is that the
transfer roughly takes place when the ground state energy of the
gate-well drops below that of the dopant-well. When the gate-well
is larger, the ground state energy is closer to the bottom of the
well and the transfer takes place at lower gate voltage.


In a realistic device, the barrier between the gate and the
semiconductor will have a finite thickness (in the most common
material systems this will be at least 1 to 2~a.u.). Usually, this
barrier does not have the same dielectric constant as the
semiconductor and hence it can modify the gate potential
considerably. Moreover, the gate must be connected to the outside
world by some kind of interconnect. Such an interconnect must be
separated from the semiconductor by a much thicker barrier in
order to sufficiently screen its potential \cite{smit02c}.
Therefore, in a realistic device, the gate must be buried in a
thick layer of barrier material.

To allow for comparison with the idealized situation in which our
calculations were carried out, several calculations were repeated
with a realistic barrier present. To that end, we obtained the
potential landscape due to the gate by solving the Poisson
equation with a finite element method (FEM) \footnote{Contrary to
the situation without a barrier where the backgate was thought to
be at infinity, in the FEM-calculations a backgate had to be put
at a finite distance from the barrier, which was chosen as 50~a.u.
in the presented calculation. The results depend only weakly on
this distance, provided it is much larger than $d$.}. It was found
that for typical realistic parameters (e.g. a SiO$_2$/Si-system
with $\es=12$, $\eb=4$, and $\dbar=2$~a.u.), the potential
landscape \emph{in the semiconductor} is qualitatively similar to
the situation where the gate is put directly on the semiconductor.
As demonstrated in the inset of Fig.~\ref{fig:hyper}, the same
phenomena are observed, but they occur at a higher voltage than in
the absence of a barrier. The voltage drop over the barrier can
roughly be accounted for by a linear scaling factor that depends
on $\eb$ and $\dbar$. Indeed, we find from the FEM-calculations
that for the given parameters about 31\% of the gate voltage drops
in the semiconductor. This number is similar to the observed ratio
between the ionization voltages with and without a finite barrier
thickness. This justifies the presentation of mainly results
obtained with an idealized barrier.

As a final remark in our discussion of the barrier, we note that
for any application or measurement of a single dopant device, it
is crucial that there are no charge traps present near the dopant.
Therefore it is highly desirable to have the barrier epitaxially
grown on the semiconductor. A promising candidate is a
Si$_{1-x}$Ge$_x$-layer as barrier on a Si substrate \cite{kane00},
although the maximum achievable barrier height in this system is
only about 100~meV \cite{walle00}.

The presented time-\emph{in}dependent calculations are not
sufficient to predict whether the dopant atom will indeed be
ionized when the ground state wave function has a low electron
density at the dopant site. In order to complete our analysis, an
estimate of the tunnel probability is needed. This is obtained by
comparison with the resonance lifetime of a hydrogen atom in an
electric field. The typical field strengths considered in the
region between the gate and the dopant site are very large (e.g.\
0.05--0.5~a.u.\ for $\rA=2$, $d=10$, $V_\mathrm{g}=2$, see
Fig.~\ref{fig:elpot}) . Using a calculation of the Stark effect in
hydrogen \cite{ivanov97} while taking the value of Ry$^*$ for
silicon, it is found that the electron lifetime at the dopant site
ranges roughly from 0.1~ps to 1~ns. This can be interpreted as the
time it takes for the dopant to be ionized when the gate voltage
is switched on and justifies our interpretation of
Fig.~\ref{fig:hyper} as the representation of an ionization
process.


Our general analysis can be readily applied, as we performed the
calculations with parameters that are consistent with the quantum
computer design mentioned. First, controlled tuning of the
hyperfine interaction by the gate, which is required in
Ref.~\onlinecite{kane98}, is possible only when $d$ is small
enough: from Fig.~\ref{fig:hyper} we estimate $d\lesssim 6$~a.u.
Switching off the hyperfine interaction, as required in the
`digital approach' \cite{skinner03}, can only be achieved for
large separation between dopant and gate ($d\gtrsim 10$~a.u.).
Hence, the dimensions of the device determine in which of both
regimes operation takes place. Second, our analysis can be used to
estimate the required gate voltage to tune the hyperfine
interaction to a certain value (Fig.~\ref{fig:hyper}). Third, it
is found that the required voltage to fully ionize the dopant
depends on $\rA$, but it is nearly independent of $d$
(Fig~\ref{fig:hyper}).


In conclusion, we analyzed the wave function manipulation of a
semiconductor dopant atom by a small electrostatic gate. We find
that two regimes can be distinguished for the ionization process
of the dopant. For a dopant-gate separation smaller than $\sim
8a_0^*$ (e.g. $\sim 24$~nm for P in Si), the electron is gradually
pulled out of the Coulomb potential of the dopant. When the dopant
resides further away from the gate, the transfer takes place
abruptly at a well-defined threshold field. Both regimes are
accessible, since, e.g., epitaxial growth techniques allow for
sufficiently accurate positioning of the dopant under the gate.

\begin{acknowledgments}
We thank J.~R.~Tucker for useful discussions. One of us, S.R.,
acknowledges the Royal Netherlands Academy of Arts and Sciences
for financial support.
\end{acknowledgments}

\end{document}